\documentclass[%
reprint,
superscriptaddress,
amsmath,amssymb,
aps,
floatfix,
]{revtex4-1}

\usepackage{graphicx}% Include figure files
\usepackage{dcolumn}% Align table columns on decimal point
\usepackage{bm}% bold math
\usepackage{graphicx}
\usepackage{amssymb}	%% Font - using for \square
\usepackage{amsmath}	%% Font - created by American Mathematical Society
\usepackage{bm}			%% Font - bold font in Italics (ex. bm{comments})
\usepackage{upgreek}	%% Font - up greek
\usepackage{mathrsfs}
\usepackage[pdflinkmargin=5pt,pdfstartview={FitBH -32768},pdfpagemode=None,plainpages=false,colorlinks=true,linkcolor=blue,bookmarks=true]{hyperref}
\begin{document}
\preprint{APS/123-QED}
%%%%%%%%%%%%%%%%%%%%%%%%%%%%%%%%%%%%%%%%%%%%%%%%%%%%%%%%%%%%%%%%%%%%%%%%%%%%%%%%%%%%%%%%%%%%%%%%%%%%
%%%%%%%%%%%%%%%%%%%%%%%%%%%%%%%%%%%%%%%%%%%%%%%%%%%%%%%%%%%%%%%%%%%%%%%%%%%%%%%%%%%%%%%%%%%%%%%%%%%%
%%%Main%%%%%%%%%%%%%%%%%%%%%%%%%%%%%%%%%%%%%%%%%%%%%%%%%%%%%%%%%%%%%%%%%%%%%%%%%%%%%%%%%%%%%%%%%%%%%
\title{Field emission from diameter-defined single-walled carbon nanotubes}

\author{Masaru Irita}
 \email{irita@rs.tus.ac.jp}
 \affiliation{Department of Physics, Tokyo University of Science, Shinjuku, Tokyo 162-8601, Japan}

\author{Yoshikazu Homma}
 %\email{homma@rs.tus.ac.jp}
 \affiliation{Department of Physics, Tokyo University of Science, Shinjuku, Tokyo 162-8601, Japan}

\begin{abstract}
Field electron emission (FE) from a structure-defined single-walled carbon nanotube (SWNT) was observed experimentally. A series of observations of the same SWNT by scanning electron microscopy, field emission microscopy and transmission electron microscopy were performed for the characterization of the SWNT emitter. This characterization work was very difficult because the observations affected the sample. Additionally, different sample setups were needed in each observation. We improved measurement conditions to minimize the influence to SWNT and developed techniques of sample setup for these observations. Combining these techniques, FE from an individual diameter-defined SWNT could be observed. The diameter of SWNT was $\sim 1.73\ \mathrm{nm}$ and the FE current of $2.2 \times 10^{-7}\ \mathrm{A}$ was obtained at a low applied voltage of $270\ \mathrm{V}$. We observed a dynamic behaviors of the SWNT emitter during FE. 
\end{abstract}

\pacs{Valid PACS appear here}% PACS, the Physics and Astronomy
                             % Classification Scheme.
                              %display desired
\maketitle

%%%%%%%%%%%%%%%%%%%%%%%%%%%%%%%%%%%%%%%%%%%%%%%%%%%%%%%%%%%%%%%%%%%%%%%%%%%%%%%%%%%%%%%%%%%%%%%%%%%%
\section{Introduction}
	Field emission microscopy (FEM) was invented by Erwin M{\"u}ller in 1936 \cite{1936_Muller_ZfPhys}. The FEM image shows surface atomic structures directly and analysis of field electron emission (FE) characteristics reveal electronic properties of the surface of tip-shaped samples. FEM is used to study electron emitters of scanning electron microscopy (SEM), transmission electron microscopy (TEM), etc. In recent years, carbon nanotubes (CNT) have been studied intensively because the unusually high aspect ratio as well as the chemical stability of CNT is the main advantage over the conventional metallic tips \cite{2010_Saito_Wiley}. Particularly, single-walled carbon nanotubes (SWNTs) are attractive nanomaterials having excellent properties theoretically predicted \cite{2002_Han_PRB}. Most of the properties have been confirmed experimentally. In the field of FEM, multi-walled carbon nanotubes (MWNTs) were mostly used as CNT emitters \cite{2000_Saito_JJAP, 2002_Oshima_PRL, 2012_Heeres_PRL}. Dean {\it et al.} demonstrated use of individual SWNTs for extracting FE \cite{2000_Dean_APL, 2001_Dean_ApplPhysLett}. They reported the effect of adsorption of molecules on the FE current from SWNTs \cite{2000_Dean_APL}. At a high current regime exceeding a Fowler-Nordheim behavior, they showed field evaporation of the atoms on the end of SWNT, resulting in reduced nanotube length \cite{2001_Dean_ApplPhysLett}. They reported a spinning motion of FEM images and current degradation due to the field evaporation. Marchand {\it et al.} reported the FE from growing individual SWNTs \cite{2009_Marchand_Nanoletters}. They discovered axial rotation of SWNTs during growth. Those works revealed extraordinary behaviors of SWNTs as field electron emitters. However, the FE properties from clean SWNTs below the evaporation regime have not been understood yet. Furthermore, although FEM images of SWNT caps have been studied \cite{2003_Dean_JVacSciTecB}, no systematic relationships between FE images and cap structures were found. So far, FE from a structure defined SWNT has not been explored yet because of the difficulty in the sample treatment. It is important to observe FE from structure-defined SWNTs to understand the FE characteristics from SWNTs. The purpose of this study is to establish the observation techniques of SWNT emitter and to observe the FE from a structure-defined SWNT.

	We have developed technologies for fabrication of SWNT tips and could produce many SWNT tips stably at the yield rate of $25\%$. In the previous work, SWNT tips was examined as the probe of ultrahigh vacuum scanning tunneling microscopy (STM) \cite{2013_MasaruIrita_eJSSN}. We found that the length of SWNT was a crucial factor for the application to STM \cite{2013_MasaruIrita_eJSSN}. For FE, the length of SWNT is important too. In the case of a SWNT tip with length of $\geq 1000\ \mathrm{nm}$, the FE current was not stable. We could not observe an FEM image because of the instability, and found disappearance of SWNT on the tip by SEM. Therefore, we needed to fabricate an isolated SWNT as short as possible on a W tip. 

	The fabrication of SWNT tip required SEM observation of the products to find an individually standing SWNT on the W tip. We selected only good samples for FEM observation among the products. Furthermore, the SWNT tip was observed by TEM and the diameter was determined. We need a series of observations of the same SWNT by SEM, FEM and TEM to understand the relationship between FEM images and the SWNT cap structure. This characterization work was very difficult because the observations affected the sample. Additionally, the sample setup was different in each observation. Success of observations depends greatly on the sample setup method. We improved techniques of sample setup for these observations. We show a few successful experimental results. 

%\clearpage
%%%%%%%%%%%%%%%%%%%%%%%%%%%%%%%%%%%%%%%%%%%%%%%%%%%%%%%%%%%%%%%%%%%%%%%%%%%%%%%%%%%%%%%%%%%%%%%%%%%%
\section{Experimental}
	In this study, we produced many W tips (length of $10\ \mathrm{mm}$ and $0.3\ \mathrm{mm}$ in diameter) simultaneously with an automated electropolishing device that had paralleled circuits of nanosecond-order switching. On the W tip, SWNT was directly synthesized by vacuum deposition of catalyst metals and chemical vapor deposition of carbon \cite{2013_MasaruIrita_eJSSN}. Fabricated SWNT tips were observed by SEM, FEM and TEM with minimized influence to the sample. The sample setup is different in each observation as shown in Fig. \ref{fig1}. SWNT tip samples were very likely to break with clumsy treatments. In particular, TEM sample preparation was very difficult and SWNT disappeared often. We developed techniques for the observations as shown in the following section. Combining these techniques, FE from an individual diameter-defined SWNT could be observed. 

%%%%%%%%%%%%%%%%%%%%
\subsection*{SEM observation}
	First, the fabricated SWNT tips were observed by SEM and the length and position of SWNT on the W tip were determined. Ten W tips were mounted on each sample stage for SEM observation as shown in Fig. \ref{fig1} (a). The fabricated SWNT tips were classified into three types as shown in Fig. \ref{fig2}. About $43\%$ of products had standing SWNTs, but about $30\%$ of SWNT tips were not usable for FEM because of off-axis SWNT growth or growth out of the tip apex. About $10\%$ of SWNT tips were longer than $1000\ \mathrm{nm}$ on the tip apex. Only $3\%$ of SWNT tips were shorter than $500\ \mathrm{nm}$ and good for this study. 
	
	The primary electron energy for SEM observation was $0.5\ \mathrm{keV}$. The electron beam did not break SWNT samples. However, the SEM observation caused deposition of electron-irradiation-induced contamination and made the diameter determination by TEM difficult. We needed to reduce the electron influence. We tried to minimize the SEM observation time to $\leq 30\ \mathrm{s}$ with a total electron dose of $\leq 2 \times 10^{17}\ \mathrm{cm^{-2}}$. In this case, the SWNT diameter could be determined by TEM observation even after SEM observation. 

%\clearpage
%%%%%%%%%%%%%%%%%%%%
\subsection*{FEM observation}
\begin{figure}[!t]
\centering \includegraphics[width=\columnwidth,keepaspectratio]{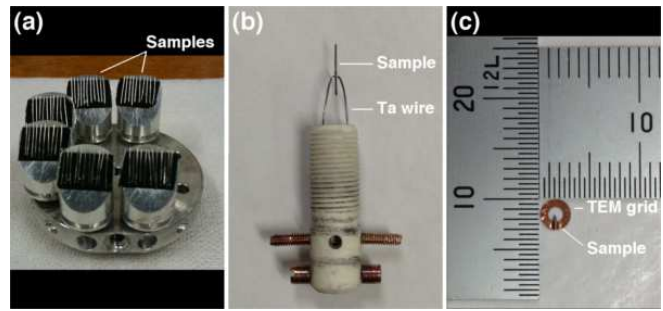}
\caption{Photographs of sample setups for 
(a) SEM, 
(b) FEM and 
(c) TEM observations. 
}
\label{fig1}
\end{figure}

\begin{figure}[!t]
\centering \includegraphics[width=\columnwidth,keepaspectratio]{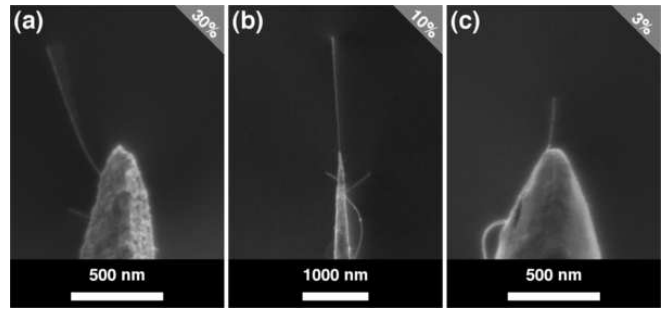}
\caption{SEM images showing typical types of SWNT tips and their abundance. 
(a) Off-axis growth: $30\%$. 
(b) Long on-axis growth: $10\%$. 
(c) Short on-axis growth: $3\%$. 
}
\label{fig2}
\end{figure}

\begin{figure}[!t]
\centering \includegraphics[width=\columnwidth,keepaspectratio]{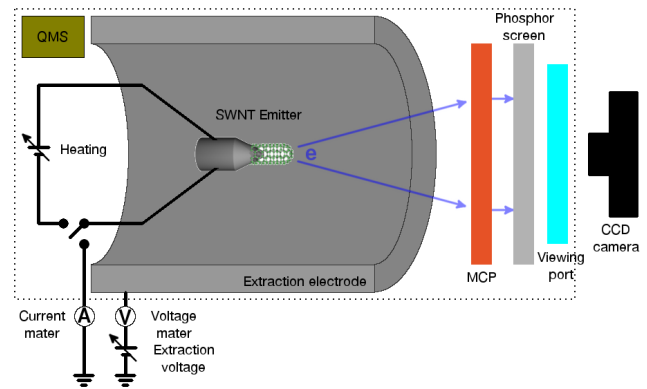}
\caption{Schematic diagram of the FEM apparatus. 
The SWNT tip is mounted inside a vacuum chamber. The voltage applied to the extraction electrode and the load current of the SWNT emitter are monitored during FE experiment. Emitted electrons are amplified through a micro-channel plate (MCP) and detected with a phosphor screen in front of the tip. FEM image on the screen is recorded with a CCD camera (the minimum illumination is $2 \times 10^{-5}\ \mathrm{Lux}$ and the image acquisition rate is $\sim 1/25\ \mathrm{s}/\text{flame}$). Gas species are analyzed with a QMS. 
}
\label{fig3}
\end{figure}

	Second, the SWNT tip was installed in a sample chamber of FEM. The W wire with SWNT tip was welded to a Ta filament as shown in Fig. \ref{fig1} (b). Figure \ref{fig3} shows a schematic of the FEM apparatus used in this study. The sample chamber was evacuated down to ultra high vacuum of $4.0 \times 10^{-8}\ \mathrm{Pa}$. To clean the surface of SWNT tips, heating was carried out to the tip up to $< 400{}^\circ\mathrm{C}$ for $5\ \mathrm{min}$. In this treatment, we took special care not to disappear the SWNT by watching gas species and their partial pressures in the mass spectra measured by a quadrupole mass spectrometer (QMS). During FEM observation, we measured the FE current $I$, applied voltage $V$ to extraction electrode, movie of FEM image and mass spectrum in the chamber at the same time. These quantities permit detailed analyses of the FE from SWNT. 


%%%%%%%%%%%%%%%%%%%%
\subsection*{TEM observation}
	Finally, after FEM observation, the SWNT tip was observed by TEM and the diameter was determined. Figure \ref{fig1} (c) shows the sample setup. In TEM observation, it is necessary to process the sample to the size of about $1\ \mathrm{mm}$ as shown in Fig. \ref{fig1} (c). This was a very difficult task and SWNT disappeared often because of impact by mechanical processing. Nevertheless, we have obtained a few successful experimental results as follows, although we need to further improve the processing method. 

	The primary electron energy for TEM was $200\ \mathrm{keV}$. The electron beam can break the SWNT sample easily. Therefore, we need to observe the sample after all of other measurements. 

%\clearpage
%%%%%%%%%%%%%%%%%%%%%%%%%%%%%%%%%%%%%%%%%%%%%%%%%%%%%%%%%%%%%%%%%%%%%%%%%%%%%%%%%%%%%%%%%%%%%%%%%%%%
\section{Results and discussion}
\begin{figure}[!t]
\centering \includegraphics[width=\columnwidth,keepaspectratio]{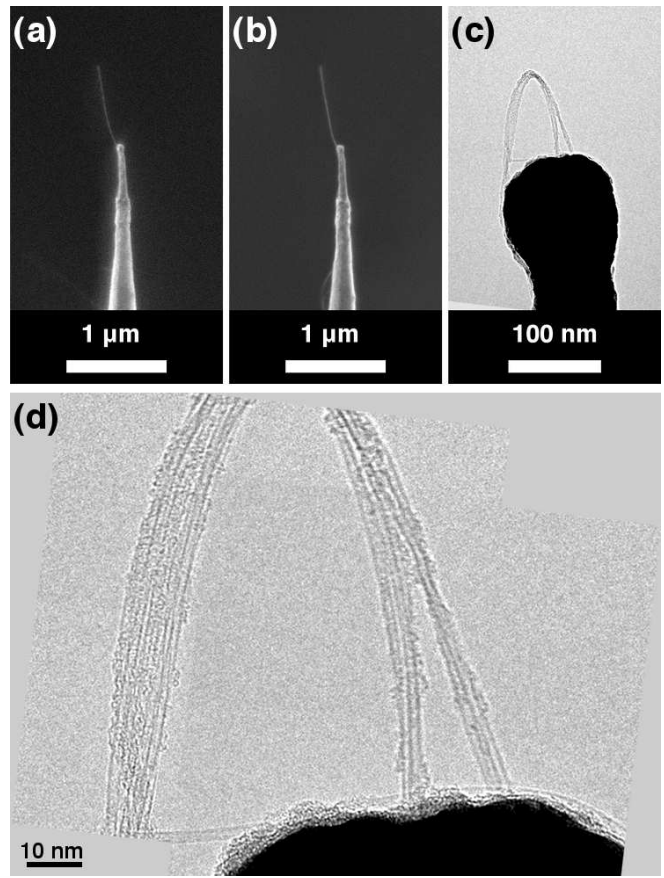}
\caption{Multifaceted observation results of the SWNT tip. 
(a, b) SEM images of the SWNT tip: 
(a) before FEM and 
(b) after FEM. 
(c, d) TEM images of the SWNT tip observed after FEM: 
(c) low magnification image and 
(d) high magnification image. 
}
\label{fig4}
\end{figure}

	Figure \ref{fig4} shows SEM and TEM images of the same SWNT tip sample. The length of SWNT was $800\ \mathrm{nm}$ as shown in Fig. \ref{fig4} (a). The SWNT tip barely changed the shape before and after FEM observation as shown in Figs. \ref{fig4} (a) and (b). After FE measurements, TEM observation was performed. However, the sample processing caused the SWNT tip bent and adhesion to the the W tip as shown in Fig. \ref{fig4} (c). Accidentally, this made stable TEM observation possible, otherwise vibration of the tip hindered observation of the tip at a high magnification. Although the very end of the tip could not be observed, the TEM image in Fig. \ref{fig4} (d) clearly shows that the SWNT tip is a four-SWNT bundle with the average SWNT diameter of $1.73\ \mathrm{nm}$. The isolated SWNT kept being extruded on the W tip until the SEM observation performed after FEM observation as shown in Fig. \ref{fig4} (b). We believe that an isolated SWNT would be extruded from the bundle of SWNTs. If the SWNT tip was processed well, the apex of SWNT tip could not have been observed easily because of vibration of the cantilever structure.

\begin{figure*}
\centering \includegraphics[width=\textwidth,keepaspectratio]{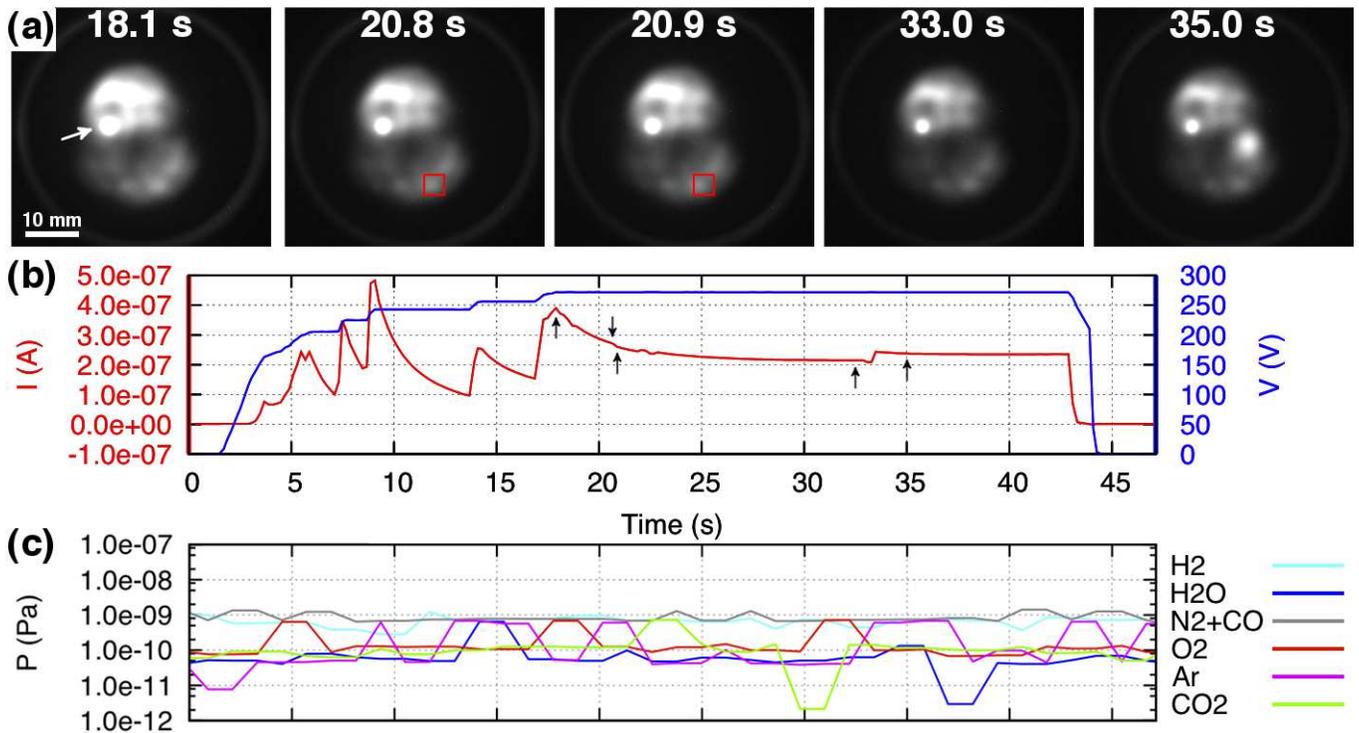}
\caption{
FEM observation results of the SWNT tip in $1.3 \times 10^{-7}\ \mathrm{Pa}$. 
(a) Snapshots of the FEM image. 
The arrow shows a defect of MCP. 
The red squares show the same position on the screen. 
The FEM image is strongly related to the FE current. 
(b) Time variation of $I$ and $V$. 
(c) Mass spectra during FEM experiment. 
The time axis is common in (b) and (c). 
The arrows in (b) correspond to each snapshot of FEM image. 
}
\label{fig5}
\end{figure*}

	Figure \ref{fig5} (a) shows the FEM images. The FE from SWNT behaved like Figs. \ref{fig5} (b)-(c). We observed these data at the same time. Each snapshot of FEM image in Fig. \ref{fig5} (a) corresponds to the circles indicated on the $I$-$t$ curve in Fig. \ref{fig5} (b).

From those results, the following three characteristics are drawn.
First, the FE occurs with a low applied voltage of $150\ \mathrm{V}$ as shown in Fig. \ref{fig5} (b).
Second, at the instant the extraction voltage was applied, the FE current was increased suddenly and  then decreased. The FE current exhibits sawtooth shapes and depends on the applied voltage as shown in Fig. \ref{fig5} (b). When the applied voltage is fixed at $270\ \mathrm{V}$, a stable FE current of $2.2 \times 10^{-7}\ \mathrm{A}$ could be obtained as shown in Fig. \ref{fig5} (b). 
Third, the FEM image of SWNT sometimes twist back and forth slightly. We observed the twisting of FEM images $\sim 1^\circ$ between $20.8\ \mathrm{s}$ and $20.9\ \mathrm{s}$ in Fig. \ref{fig5} (a). However, the twisting cannot be recognized easily from the snapshot images because of a small twisting angle. It can be recognized easier in the movie (see supporting movie). This phenomenon was reflected to the FE current. The FE current change with twisting was $0.1 \times 10^{-7}\ \mathrm{A}$ at $270\ \mathrm{V}$ as shown in Fig. \ref{fig5} (b). This phenomenon is different from adsorption of gas molecule between $33.0\ \mathrm{s}$ and $35.0\ \mathrm{s}$ in Fig. \ref{fig5} (a). The FE current change with gas adsorption was $0.2 \times 10^{-7}\ \mathrm{A}$ at $270\ \mathrm{V}$, larger than that of twisting. The twisting angle was small and the period was $\sim 1/25\ \mathrm{s}$. Furthermore, we did not observe any gas peaks of carbon-bearing molecules induced by FE as shown in Fig. \ref{fig5} (c). We have observed various SWNT tips, and observed similar twisting. At present, the exact reason of the FEM image twisting is not clear. The observed motion is different from the spinning motion of FEM images due to the field evaporation \cite{2001_Dean_ApplPhysLett} or "screw-dislocation-like" growth mechanism \cite{2009_Marchand_Nanoletters}. We speculate that a twist motion of the SWNT bundle could be the origin of it.

	Here, we should comment on the transient behavior of the FE current upon increase in the applied voltage (Fig. \ref{fig5} (b)). Dean {\it et al.} reported rapid FE current decays coincided with the field evaporation of SWNT at high FE currents \cite{2001_Dean_ApplPhysLett} and adsorption in non-ideal vacuum conditions \cite{2000_Dean_APL}. However, our FEM observation condition was different from their environments. The FE current and applied voltage were much lower and did not induce the evaporation of SWNT. The FEM images did not change in the shape during the FE processes as shown in Fig. \ref{fig5} (a). Observed FE current change was reproducible unless the SWNT emitter was broken. We assume that the phenomenon is related to transient change in the tip temperature due to Joule heating by FE current, but details are under investigation.

%%%%%%%%%%%%%%%%%%%%%%%%%%%%%%%%%%%%%%%%%%%%%%%%%%%%%%%%%%%%%%%%%%%%%%%%%%%%%%%%%%%%%%%%%%%%%%%%%%%%
\section{Conclusions}
	We have developed a multifaceted characterization method for SWNT tips, and succeeded in measuring field electron emission properties and observing FEM images from a diameter-defined SWNT. We showed dynamic behaviors of the SWNT tip during FE and confirmed the advantage of low applied voltage to obtain FE. We hope to determine the cap structure of an SWNT from FEM images, by accumulating FEM images from diameter-defined SWNTs. 

%%%%%%%%%%%%%%%%%%%%%%%%%%%%%%%%%%%%%%%%%%%%%%%%%%%%%%%%%%%%%%%%%%%%%%%%%%%%%%%%%%%%%%%%%%%%%%%%%%%%
\bibliographystyle{apsrev4-1.bst}
\bibliography{library.bib}
%%%%%%%%%%%%%%%%%%%%%%%%%%%%%%%%%%%%%%%%%%%%%%%%%%%%%%%%%%%%%%%%%%%%%%%%%%%%%%%%%%%%%%%%%%%%%%%%%%%%

\end{document}